\documentstyle[11pt,epsf,graphics]{article}

\renewcommand{\arraystretch}{1.3}

\makeatletter
\newdimen\normalarrayskip              
\newdimen\minarrayskip                 
\normalarrayskip\baselineskip
\minarrayskip\jot
\newif\ifold             \oldtrue            \def\new{\oldfalse}
\def\arraymode{\ifold\relax\else\displaystyle\fi} 
\def\eqnumphantom{\phantom{(\theequation)}}     
\def\@arrayskip{\ifold\baselineskip\z@\lineskip\z@
     \else
     \baselineskip\minarrayskip\lineskip2\minarrayskip\fi}
\def\@arrayclassz{\ifcase \@lastchclass \@acolampacol \or
\@ampacol \or \or \or \@addamp \or
   \@acolampacol \or \@firstampfalse \@acol \fi
\edef\@preamble{\@preamble
  \ifcase \@chnum
     \hfil$\relax\arraymode\@sharp$\hfil
     \or $\relax\arraymode\@sharp$\hfil
     \or \hfil$\relax\arraymode\@sharp$\fi}}
\def\@array[#1]#2{\setbox\@arstrutbox=\hbox{\vrule
     height\arraystretch \ht\strutbox
     depth\arraystretch \dp\strutbox
     width\z@}\@mkpream{#2}\edef\@preamble{\halign
\noexpand\@halignto
\bgroup \tabskip\z@ \@arstrut \@preamble \tabskip\z@ \cr}%
\let\@startpbox\@@startpbox \let\@endpbox\@@endpbox
  \if #1t\vtop \else \if#1b\vbox \else \vcenter \fi\fi
  \bgroup \let\par\relax
  \let\@sharp##\let\protect\relax
  \@arrayskip\@preamble}
\def\eqnarray{\stepcounter{equation}%
              \let\@currentlabel=\theequation
              \global\@eqnswtrue
              \global\@eqcnt\z@
              \tabskip\@centering
              \let\\=\@eqncr
 \halign to \displaywidth\bgroup
    \eqnumphantom\@eqnsel\hskip\@centering
    $\displaystyle \tabskip\z@ {##}$%
    \global\@eqcnt\@ne \hskip 2\arraycolsep
         $\displaystyle\arraymode{##}$\hfil
    \global\@eqcnt\tw@ \hskip 2\arraycolsep
         $\displaystyle\tabskip\z@{##}$\hfil
         \tabskip\@centering
    &{##}\tabskip\z@\cr}
\begingroup\ifx\undefined\newsymbol \else\def\input#1 {\endgroup}\fi
\newfont{\hr}{msbm10}
\newfont{\ams}{msam10}

\hoffset= - 1.0in

\def\beq{\begin{equation}}
\def\eeq{\end{equation}}
\def\ba{\beq\new\begin{array}{c}}
\def\ea{\end{array}\eeq}
\def\be{\ba}
\def\ee{\ea}

\def\N2{${\cal N}=2$}

\def\1N{${\cal N}=1$}
\def\4N{${\cal N}=4$}

\def\mtm{MTM}
\def\mbh{MBH}

\title{{\bf
If LHC is
a Mini-Time-Machines Factory,\\
Can We Notice?\footnote{Published in the Proceedings of the Balkan
Workshop 2005, 19-23 May 2005, Vrnjacka Banja, Serbia: Facta
Universitatis (Series: Physics, Chemistry and Technology), {\bf 4}
(2006) 381-404}} \vspace{0.cm}}
\author{{\bf A.Mironov}\thanks{E-mail:
\ mironov@itep.ru; mironov@lpi.ac.ru}
\date{ } \\
{\small {\it Theory Department, Lebedev Physics Institute}
and {\it ITEP, Moscow, Russia}}\\
{\bf A.Morozov}\thanks{E-mail: \ morozov@itep.ru}
\date{ } \\ {\small
{\it ITEP, Moscow, Russia}
}\\
{\bf T.N.Tomaras}\thanks{E-mail:
\ tomaras@physics.uoc.gr}
\date{ } \\
{\small {\it Department of Physics and Institute of Plasma Physics, University of Crete,
and Fo.R.T.H., Greece}}}

\begin{document}

\setcounter{footnote}{3}

\maketitle

\vspace{-11.5cm}

\begin{center}
\hfill FIAN/TD-12/05\\
\hfill ITEP/TH-51/05\\
\end{center}

\vspace{8.5cm}

\begin{abstract}
Assuming the hypothesis of TeV-scale multi-dimensional gravity,
one can imagine that at LHC not only mini-black-holes (MBH)
will be intensively created, but also other exotic gravitational
configurations, including hypothetical mini-time-machines (MTM).
Like MBH, they should quickly evaporate,
but one can wonder if their temporal existence at the moment of
high-energy collision can leave any traces in the observable data.
We briefly discuss five thinkable effects:

(i) change of the energy spectrum
due to the frequency-filtration property of MTM,

(ii) possible production of anomalously energetic particles,
accelerated by passing many times through gravitational field inside
the MTM,

(iii) acceleration of particle decays, since the proper time
of a particle moving inside MTM can strongly exceed the laboratory time,

(iv) CPT and naive unitarity violation (thermalization)
due to effective non-local
interactions caused by MTM and to possible ambiguity in the
population of closed world-lines inside MTM,

(v) collective effects due to conversion of a single particle into
a bunch of its co-existing copies within the MTM.

Despite possible particle-antiparticle conversion inside MTM, they
do not seem to produce any specific CP-violation effects.
\end{abstract}

\newpage

\section{Introduction}

The string-theory-inspired TeV-Gravity models \cite{TVG},
where matter lives on a 3-brane embedded into a
$D$-dimensional space-time with the extra $D-4$ dimensions
compactified on a manifold of the inverse size $\sim 1\div 10 {\rm TeV}$,
opened a potential possibility to observe non-trivial
gravitational effects in accelerator \cite{acc},
cosmic ray \cite{mmt} and neutrino \cite{nuMBH} experiments.
So far, attention in this field was restricted to effects, associated with
the possible production of mini-black-holes (MBH), which
can be {\it massively} produced already at LHC if TeV-Gravity models
are true. These MBH are supposed to evaporate
instantaneously (see, however, \cite{Fron}),
but -- since Hawking radiation does not
distinguish between the sorts of particles and space directions --
can cause energy and momentum redistribution between the products
of reaction, which is potentially observable (though not
too pronounced -- as usual for the TeV-energy experiments).\footnote{We need
to mention here that while discussing experimental evidence of
the black hole evaporation one has to look after various conservation
laws like baryon number conservation. In particular, we believe that the
probability of black hole formation by quarks within a proton with a
subsequent evaporation, i.e. a proton decay into mesons and, possibly,
leptons is vanishing. Basing on some arbitrary assumptions about what
quantum gravity is, \cite{protonHaw}, one may estimate \cite{proton2}
the proton lifetime to be $M_{Pl}^{d+4}/m_p^{d+5}$ ($4+d$ is the number of
dimensions where quarks propagate) which is unacceptable large for
Tev-Gravity models. Note, however, that this is completely {\it quantum}
gravity (whatever it means) effect, while forming black holes in
accelerators is completely {\it classical} and, therefore, say, arguments in
\cite{Stoj} are absolutely misleading.}

It is natural to assume that the story should not be restricted
to MBH: once Pandora box
of gravitational effects on particle physics is open, all of its
content can show up. This implies that other types of
multi-dimensional gravitational configurations could also be born,
along with the MBH. Since much less is known about
such configurations, except that they appear on equal footing
with black holes in general relativity, we can just {\it assume}
that the same is true in particle theory: namely,
that, like MBH, they are created in particle collisions
{\it classically}, with the cross-sections,
defined by their geometrical sizes, with only a modest damping caused
by radiation of gravitational waves, and decay almost instantly,
either classically or due to some analogue of Hawking radiation.
If one agrees to accept this assumption,
the question arises whether any such configurations can leave traces
in the properties of particle collisions,
essentially different from those of mini-black-holes.

The most interesting from this perspective are the
(mini-)black-\-rings \cite{br} and, especially,
(mini-)time-\-machines (MTM): geometries with existing closed time-like
worldlines. There is a long story of discussions around
time-machines in general relativity \cite{tm}, we agree with
the recent conclusion of \cite{viss,Ori,kr1},
that no reasons were found to forbid their existence.\footnote{
Touching the subject of time machines, one, probably, should not avoid
comment on the celebrated grandfather, known also as butterfly or,
in scientific literature, as
Polchinski \cite{Polch}, paradox: what if, while traveling to the
past, one affects or even destroys the preconditions for its own creation.
Here one should be careful in distinguishing between ``affects" and ``destroys".
In the {\it former} case, there is no trouble: one should just modify the
relation between initial and final conditions (causes and corollaries),
remembering also that solutions to differential equations on spaces
with non-trivial topology are uniquely defined by initial conditions
for non-compact directions {\it and} by ``zero-modes" for compact ones (in other
words, the ``residents" of the time-machine should be included into
formulation of the Cauchy problem). In the {\it latter} case,
the ``problem" is actually a one of a G\"odel- or Echer-style
``impossible things": describing in a Human-created language
a non-existing entity. In quantum theory, one should sum over
all possible {\it globally defined} histories (world lines), and the
paradox normally describes in words a (self-contradictory) history,
which actually does not exist:
if {\it somebody} killed your grandfather, then
his world line terminated and could not lead to your birth-day, thus,
that {\it somebody} was not you. Hence, there is no continuous world line,
satisfying the conditions of the paradox, and the
{\it non-realizable} history of {\it you}
killing {\it your} grandfather does not
contribute to the functional integral. It is kinematically,
not dynamically, forbidden. Thus, the scientific problem related to grandfather
paradox is that of classical determinism and of accurate formulation of
Cauchy-like problems, existence and uniqueness of its solution,
in the presence of time machines. It is sometime a difficult, but in no
way a paradoxical problem. See \cite{gfp} for more considerations.}
Thus we suggest to switch the direction of the
discussion: from whether time-machines are allowed to exist
and be observable
(not separated from us by any sort of impenetrable horizon like
in the case of non-traversable wormhole, say, the Einstein-Rosen
bridge, \cite{Wheeler}), to whether we can notice them.
As explained above, the TeV-gravity models allow one to shift
this discussion to the solid ground of particle physics and
perhaps even to forthcoming accelerator experiments.
In this context the basic question is: can massively produced and
quickly ``evaporating" mini-time-machines
leave traces, somehow different from those of pure-particle-physics
processes (with no gravity effects) and from those caused by
mini-black-holes?
We do not say that discussion of mini-time-machines production rates
are not important -- quite the opposite, we shall see that these
rates can essentially affect the answer and need be evaluated, --
just, given all the experience of discussion of
``chronology protection principle" \cite{Haw}, we believe that one can
try to overstep this controversial subject and see what happens:
if something interesting occurs, then we can come back with more
enthusiasm and devotion.

One more word of caution is needed, this time not from the
general-relativity, but from the particle-theory side.
The best one can technically do at the moment is to consider MBH and MTM,
once they are created, as classical backgrounds and see what
happens to particles embedded into them. In MBH case it was
sufficient to accept the existence of Hawking radiation,
in the MTM case we discuss below a some more delicate
aspects of particle propagation. Strictly speaking, such consideration
is safe when the scale of the background exceeds the particle's
Compton wavelength -- what is hardly true for
MBH and MTM, created in high-energy collisions (though one of the
two parameters of the MTM -- duration $T$ -- can actually be large).
We {\it assume} that qualitative effects can still be evaluated
in above ``approximation", like it is assumed about the Hawking
radiation of MBH, but more justification is needed here.

Surprisingly small is known about the testable properties of
time machines: most discussions seem to concentrate on philosophical
issues and not much is done in estimation (and even definition)
of concrete quantities.
In the rest of this paper, we try to list possible qualitative effects
in particle physics that could be of interest for further
quantitative investigation.

\section{Basic assumptions \label{s2}}

\subsection{On MTM-induced corrections to scattering probabilities}

In quantum theory of fields $\phi(t,\vec x)$ the mini-time-machines
modify probabilities of scattering processes in the following way:
\be
\sum_{\rm over\ geometries}
\rho_{\rm geometry}
\left|\big< \phi(t_1,\vec x_1)\ldots \phi(t_n,\vec x_n)
\big>_{\rm geometry}\right|^2
\label{f0}
\ee
This formula 
reflects the fact that the story is about {\it classical}
gravity: only matter fields are quantum, and no interference is
considered between different geometries, say, between mini-time-machines
with different duration of time loops.

\subsection{On probability of MTM creation in particle collisions}

The probability $\rho_{\rm geometry}$ of given geometry to contribute
characterizes the probability to the corresponding object
(mini-\-black-\-hole, mini-\-black-\-ring, mini-\-time-\-machine etc)
to be created at accelerator (or in cosmic ray event) at energy ${\cal E}$.
Like in the mini-\-black-\-hole case we assume that this probability is
basically defined by the geometrical size ${\cal R}$ of the object
(Schwarzschild radius or a size of the time-machine mouth) and the threshold
energy ${\cal E}_{threshold}$:
\be
\rho_{\rm geometry} \sim
{\cal R}^2\theta\left({\cal E} - {\cal E}_{threshold}({\cal R})\right),
\label{crepro}
\ee
where $\theta(x)$ is the Heaviside step-function. Of course this is a
disputable formula (``chronology protection principle", if true, would simply
put $\rho$ for time-machines equal zero, and even the analogue of this
formula for mini-black-holes caused long discussion
\cite{mbhcs}), but we just state
that at the moment nothing forbids one to {\it assume} that (\ref{crepro})
{\it can} be adequate. It just reflects the assumption that mini-time-machines
can be created {\it classically} in particle collisions,
with no quantum damping (which would characterize {\it quantum} creation of a
{\it coherent} classical, i.e. consisting of {\it many} particles,
object from {\it a few} colliding particles) and the obvious existence of some
minimal energy needed for strong deformations of flat space-time to occur.
As known from considerations of {\mbh}, the cross-section (\ref{crepro}) with
${\cal R}$ and ${\cal E}_{threshold}({\cal R}) \sim {\cal R}m^2_{Pl}$,
allowed by TeV-gravity models (i.e. with $m_{Pl} \sim 1\div 10\ {\rm TeV}$),
is enough to make LHC a {\it factory}, producing an {\mbh}\ every second,
and we can imagine that the rate of {\mtm}\ production can be comparably high.
This feeling that time machines are often associated with the black holes,
so that the probabilities of their existence can be indeed comparable,
both in the Universe and in accelerator processes, starts to get some support
in the general-relativity literature of the last years \cite{tmbh}.
Of course, there is a more delicate dependence on the parameters of \mtm\
 in the proportionality coefficient, implied in (\ref{crepro}), and it is
important for any kind of numerical and even qualitative estimates, but it
is hard to evaluate without addressing concrete models of \mtm.

\subsection{\label{mtmg} On MTM geometries}

A typical example of time-machine geometry is
obtained by cutting two $4d$ balls out of Minkowski space and identifying
(gluing) the boundaries of emerged holes (``mouths"). If one of the balls is inside
the light-cone of another, we obtain a time-machine Wheeler's wormhole, see Fig.1
(otherwise it would be a Wheeler's wormhole with no closed time-like curves).
This particular construction of the space with closed time-like curves
was originally proposed in \cite{Deu,Pol1}.
As seen from Fig.1 the MTM has two essentially different parameters:
the size of the mouth (length of the cut) ${\cal R}$ and duration
(distance between
the cuts) $T$. Note that in (\ref{crepro}) we assume that in the first
approximation the probability of MTM creation does not depend on $T$, in
particular large $T$ may not be damped as strongly as large ${\cal R}$, and
$T$ can essentially exceed
${\cal R} \sim {\cal E}/m_{Pl}^2
$.
This is a disputable assumption, but it is supported by some studies of
time machines in general relativity, implying that $T$ is indeed a soft
(nearly zero-) mode (even negative in same cases) of the time machine solutions
\cite{tzm,tmbh}.

Another popular
construction of the time-machine is a traversable wormhole suggested in
\cite{Th}, see also \cite{whole}. Important addition to Fig.1 is a tube,
connecting the balls boundaries (instead of gluing them directly,
as in Fig.1), and appearance of a new parameter: time-loop duration $T$
can generically exceed the time-machine time-life $\cal T$.
On a general class of wormhole geometries, see \cite{genwh}; on the stability
of wormholes, see \cite{morewh}; on the gravitating matter creating wormholes, see
\cite{kdv}; on rotating wormholes solutions which are presumably more stable, see
\cite{rwh}.
Other constructions of time machines can be found in \cite{tm}.
One can think about other ways to construct time-machine geometries.
We do not discuss classification of time-machine
geometries and their differences, example of Fig.1,
sometime with addition of additional parameter $T \geq {\cal T}$  will be sufficient
for our purpose of listing some interesting phenomena in particle physics,
induced by the presence of time-machines.

\begin{figure}[ht]
\epsfxsize 450pt
\begin{center}
\epsffile{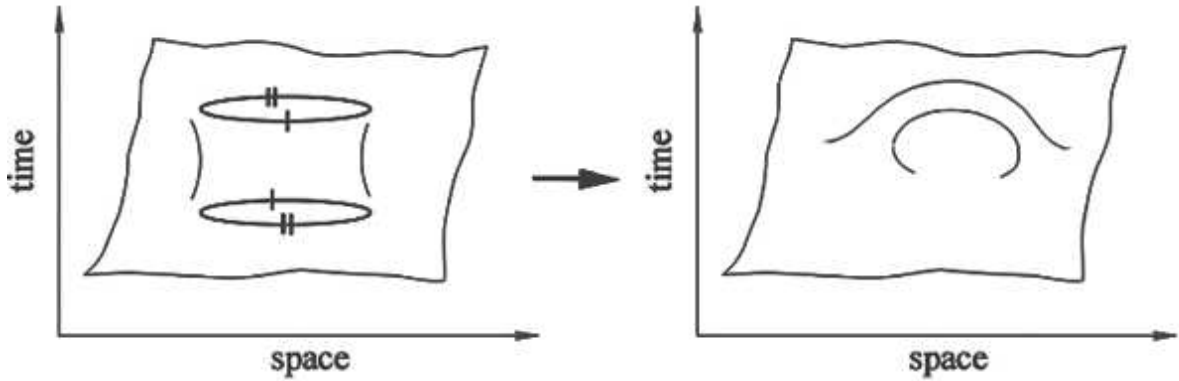}
\end{center}
\caption{\footnotesize A time-machine wormhole, represented by flat space
with cuts (with their sides appropriately identified).
The second picture shows topology of emerged space
(geometrically it can still be flat almost everywhere).
The tube in the second picture can easily change its width
far from its ends, and this allows to distinguish between the
characteristic time loop duration $T$ and the time-machine life-time
(in laboratory frame) ${\cal T}$. Both $T$ and ${\cal T}$ can be
soft modes and do not need to be restricted by collision energy as
strongly as the cut length (ball size) ${\cal R}$.}
\end{figure}

\subsection{On MTM evaporation (decay)}

If MTM contains strong gravitational fields and gravitational horizons,
they can lose energy by a direct analogue of the Hawking radiation
\cite{Kraeva}.
Remarkably, the same can be true if gravitational fields are small and no
event horizons\footnote{
Of course, a Cauchy horizon, separating the domain where initial
conditions at remote past
are not enough to fix the solutions of evolution equations
uniquely, is obligatory present in any time machine.
} are formed,
like in our flat-space-time wormhole in Fig.1.
Then time-machine's energy is fully carried by its "residents" --
particles, which move along closed world lines (or zero-modes of fields
in another formulation). Virtual-pair creation can produce an antiparticle
to annihilate any given resident, then the other element of the pair can
escape out of the time machine, thus forcing it to loose its residents
and thus the energy.
This process is possible because the finite size of the time machine
makes -- through quantum zero fluctuations -- the particles inside,
even in their classically stable ground states,
slightly more energetic than outside, what favors their escape,
whenever possible.
Properties of such evaporation should be, indeed, similar to the Hawking
radiation, in particular, it should also be thermal because of
uncertainty in the non-controllable state of the residents.

\subsection{On second quantization formalism: the key one for particle-theory
causality and unitarity}

Individual correlators in (\ref{f0}) can be represented in
two essentially different ways: in second and first quantization formalisms.
In the former case the correlator is given by a functional integral over
fields,
\be
\int D\phi\ e^{iS\{\phi\}} \phi(t_1,\vec x_1)\ldots \phi(t_n,\vec x_n),
\label{f1}
\ee
and Feynman diagrams are made out of propagators, which are obtained
by solving Klein-Gordon and Dirac equations in the space with non-trivial
topology, like in Fig.1.
It is a non-trivial problem to find explicit expressions for
such functions, as everybody knows from the study of oversimplified
examples in electrostatics or in the theory of Riemann surfaces.
Even for Fig.1, and even for the case when spheres are deformed into
straight cuts and even in $2d$ instead of $4d$ it is quite a problem to
write an adequate Green function explicitly.\footnote{In $2d$ one
readily writes down a {\it Euclidean} Green function
$$\log \left(\vartheta(\Big|z-z'\Big|\tau)|^2
\exp\pi\frac{{\rm Im}(z-z')^2}{{\rm Im}\ \tau} \right) + {\rm
contribution\ of\ the\ {zero-mode} }$$ with $\tau$ in the argument
of Jacobi theta-function defined through the ratio $T/{\cal R}$ and
$$
z\equiv\int^x \frac{dx}{\sqrt{\left(x^2-{R^2+T^2\over 4}\right)^2+T^2x^2}}
$$
is obtained by the Jacobi map.
However, this formula actually describes a Green function after
compactification of a plane to $CP^1$ and can not be directly used
for analytical continuation into non-compact Minkowski space with appropriate
boundary conditions. } If known, such expressions could be used to
somehow extract the commutators\footnote{For example, the
equal-time commutator for $t'=t$ can be found with the help of the
BJL theorem \cite{BDL}: picking up the coefficient in front of
$\omega^{-1}$ after Fourier transformation in time direction.
However, even in this case there can be problems with making Fourier
transformation in time-machine geometry, where the time $t$ is not
globally defined.} $\Big< [\phi(x,t),
\phi(x',t')]\Big>$ and check if they vanish outside the light-cone
-- and thus study the naive particle-physics causality in the
time-machine geometry. Unfortunately, because of problems with
explicit expressions, not much can be done with the
causality-related problems in the second quantization formalism.

\subsection{On first-quantization formalism}

The first-quantization formalism is much better suited to study
(counter-)intuitive problems. It represents propagators as sums over
all {\it possible} world-lines of a particle in the space-time, leading from
initial to the final point. Interactions are taken into account by
forming usual Feynman diagrams out of these propagators.
A non-trivial problem, however, arises with what are ``possible"
world-lines in Minkowski space-time.
In non-relativistic quantum mechanics \cite{FH} the sum is over all world
lines of the type $\vec x(t)$ with $-\infty < x_j(t) < +\infty$ and
$t_i< t < t_f$ and $\vec x(t_i) = \vec x_i$, $\vec x(t_f) = \vec x_f$,
i.e. one allows classical equations of motion to be broken, but
time-ordering is preserved.
In relativistic situation where the typical particle action is
$S\{\vec x(t)\} = m\int \sqrt{1 - \dot{\vec x}^2}dt$, one can impose
additional constraint $\left|\dot{\vec x}\right| \leq 1$ -- particles
can not travel faster than light. The sum over world-lines with
this restriction would provide the ``causal particle propagator"
\be
g(\vec x_f,t_f|\vec x_i,t_i) =
\int \left\{D\vec x(t)
\prod_{t=t_i}^{t_f} \theta\left(1-\dot{\vec x}^2\right)\right\}
\exp\left(
im \int_{t_i}^{t_f} \sqrt{1 - \dot{\vec x}^2}dt\right)
\label{cpp}
\ee
-- an obviously causal quantity in Minkowski space.
However, for geometrical intuition 
this object does not look very natural if space and time are to be treated
on equal footing. From that point of view, instead of  summing over all
functions $\vec x(t)$ in (\ref{cpp}) one would rather sum over all
curves $\vec x(s), t(s)$ in the space time. The difference is two-fold,
see Fig.2:
first, not all curves are inverse images of projections onto the time
segments $(t_i,t_f)$; second, not all curves satisfy the restriction
that velocities never exceed unity (the speed of light).
\begin{figure}[ht]
\epsfxsize 350pt
\begin{center}
\epsffile{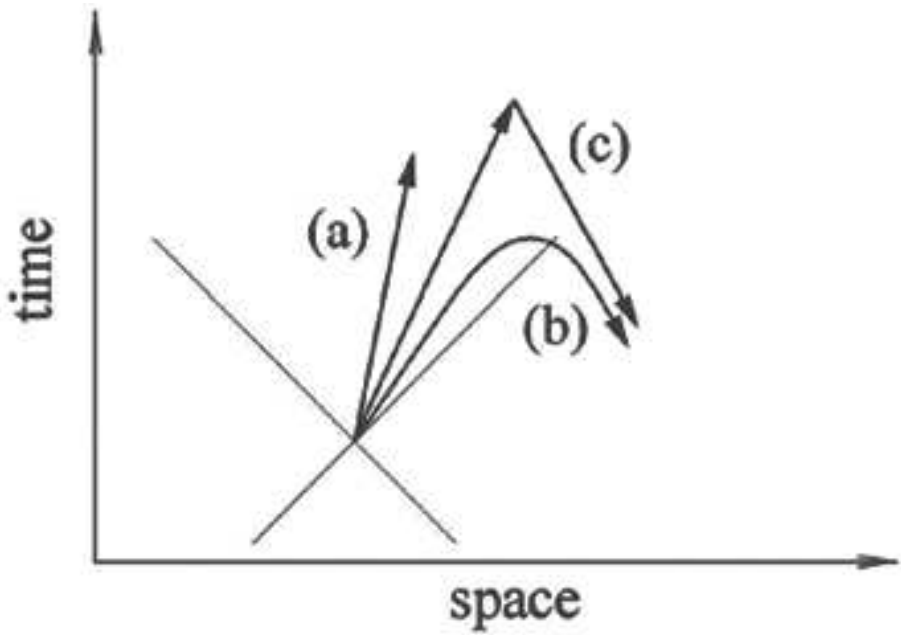}
\end{center}
\caption{\footnotesize
Difference between a particle world line (a), generic path (b)
and a singular path (c), implied in the definition of Feynman propagator.
The last one consists of pieces, which look like particle/antiparticle
world lines, i.e. belong everywhere to the light cones,
are invertibly projected on the time axis and projection
preserves/inverts the direction.}
\end{figure}
The latter subject is in fact a delicate one, related to analytical
continuations and other peculiarities of Polyakov's method
to handle non-polynomial actions \cite{Poly},
while the former difference is taken into account by
composition of any space-time curve from a sequence of well-projectable
fragments, going along and backwards in time. The full integral over
all curves -- the Feynman's propagator $G(\vec x_f,t_f|\vec x_i,t_i)$ --
is a multi-linear  combination of $g$ and $\bar g$, interpreted as
propagators of particles and anti-particles \cite{FAP}, depending on
the sign of $dt/ds$:
\be
G(\vec x_f,t_f|\vec x_i,t_i)=g(\vec x_f,t_f|\vec x_i,t_i)+
\int_{|\vec x_f-\vec x|\le t_f-t,\ |\vec x_i-\vec x|\le t-t_i}
\bar g(\vec x_f,t_f|\vec x,t) g(\vec x,t|\vec x_i,t_i)+\ldots
\ee

\begin{figure}[ht]
\epsfxsize 300pt
\begin{center}
\epsffile{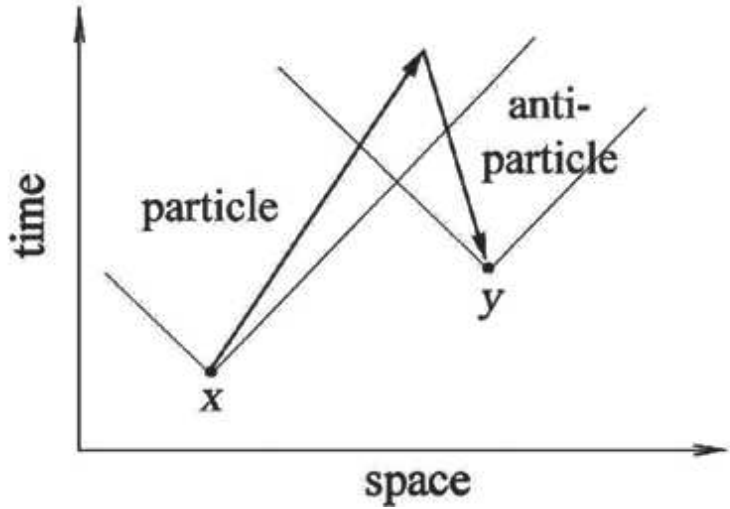}
\end{center}
\caption{\footnotesize A typical contribution to the Feynman "propagator"
outside the light-cone. Both particle and antiparticle travel inside
their light cones and no causality violation takes place. However, this
{\it correlator} does not describe {\it propagation} of any physical entity:
it is instead an amplitude to create and annihilate a pair.
This obvious fact makes interpretation of amplitudes (especially, the loop
diagrams), evaluated with the help of Feynman propagators somewhat tricky.
}
\end{figure}

As clear from Fig.3, the Feynman ``propagator" does not
vanish outside the light-cone (still, and
somewhat ironical, it is often called ``causal"), and does not need to vanish,
see, e.g., \cite{props}. Better understanding of
these problems seem important for clarification of physics of time machines,
and they seem to remain under-investigated.

\section{Particles in time-machine}

The typical world-line of a particle, traveling through time machine
of Fig.1 is shown in Fig.4.

\begin{figure}[ht]
\epsfxsize 450pt
\begin{center}
\epsffile{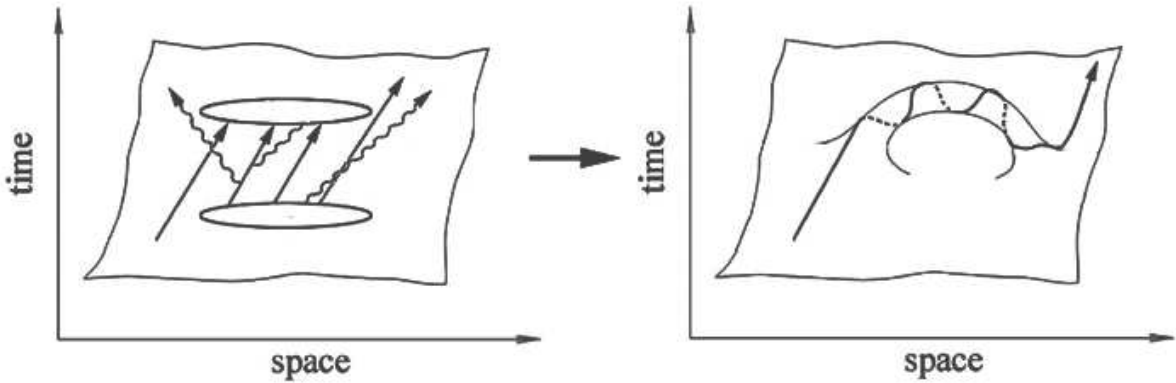}
\end{center}
\caption{\footnotesize A particle, traveling $n=3$ times through a time machine
with the time loop of length $T$. In the left-side picture also two
photons are shown, radiated by the particle. The one which escapes
immediately after being emitted looks for a distant observer as coming
from the place where no particle was present and can cause
non-locality of effective theory.}
\end{figure}

It passes $n$ times through the time-machine and finally escapes.
In laboratory frame all $n$ walks do not take any time, but in the
particle's own frame the situation is different: $nT$ units of its
proper time pass before it escapes. If the particle could decay or
radiate (in Fig.4 the radiation case is shown), it was doing so
during all its life in time-machine and comes out much older
-- by $nT$ -- than its twin particle, which never was in the time
machine (note, that
in special relativity, the traveling twin returns younger than the one
who stayed at home, but in the case of time machine, as well as of
a travel in strong gravitational field, things are different:
the traveler gets older). We now list some effects obvious from Fig.4,
together with their possible implications for observations.
In every case it is important to figure out, whether an
effect can occur

-- without any gravitational effects,

-- also in the presence of {\mbh},

-- in presence of any kind of wormholes (i.e. effect does not distinguish between
space- and time-like wormholes),

-- only for time-machine geometries.

\subsection{Frequency filtration}

The field, describing a particle with frequency $\omega$, acquires
a phase factor $e^{i\omega T}$ every time it makes a cycle along
the time loop of duration $T$ in time machine. After $n$ cycles
the wave function will be proportional to
$$
\sum_{k=0}^n e^{ik\omega T}
\longrightarrow \sum_m \delta\left(\omega - \frac{2\pi m}{T}\right)
$$
for large enough $n$. Of course, one should take into account the
space-dependence of the wave function, but the phenomenon is already
clear from above oversimplified formula: original frequency spectrum
at the entrance into the time-machine is re-shaped, mostly
frequencies which are integer multiples of $\frac{2\pi}{T}$ will
penetrate through the time machine. This frequency filtration
implies that gravitational effects can modify the original
spectra of particles, calculated in neglect of them. Actually, such
modifications are caused both by {\mtm}s and {\mbh}s. Though
effect of particular \mtm\ is quite specific: frequency filtration
(and nothing equally specific would be caused by particular \mbh),
it is obscured by averaging over various time-loop durations $T$ in
formula (1).

\subsection{Unexpectedly energetic particles}

It was noted in \cite{Haw} that the gravitational field
in a time machine could provide a particle\footnote{In \cite{Haw} it was
a photon traveling along the first closed null-curve to come into existence
as the time-machine is formed. This photon is blue-shifted every time
it makes a closed loop inside the time machine. See also \cite{Kras}.
}
with additional energy each time it passes through the time machine,
so that it can acquire a lot during a period of time which is nearly
zero in laboratory frame. This would
cause another peculiar type of modification of spectrum,
providing some particles with a large excess of energy at expense
of the other particles:
the latter are produced in evaporation of the MTM, which
lost part of its gravitational energy to acceleration of the former.
This effect is again specific
for time machines (as compared to the \mbh\ and space-like wormholes),
but again, since
spectrum modification involves a lot of averaging, it can be not
too pronounced in actual experiments.

As usual, one should add a word of caution.
Already in \cite{Haw} it was argued, that the time machines
with accelerating ability can be not observable for one or another reason
(they can decouple from our universe or instead quickly renormalize the
accelerating gravitational field down to zero).
Furthermore, in \cite{Haw} it was noted that the geometry in a tube region
around the closed curved acts as ``a diverging lens" \cite{viss},
which disperses {\it energy}, even if it concentrates the {\it energy density}.
Thus, if a point particle is substituted by a wave packet, its total
energy can actually decrease instead of increasing.
According to \cite{Haw} this is indeed the case for the wormholes of \cite{Th}.
A more careful analysis for (quantum) point-like particles is still needed.

\subsection{Accelerated aging, intensification of decay and oscillation
 processes}

If a particle travels $n$ times along a time loop of duration $T$, its
proper-time (age) exceeds the laboratory time by $nT$.
This means that from the point of view of laboratory observer such
particle ages much faster than it would in the absence of time machine.
In particular, if the particle decays and probability of its existence
decreases as $e^{-\Gamma t}$, the presence of time-machine seemingly
accelerates the decay for the outside observer:
$e^{-\Gamma t} \longrightarrow e^{-\Gamma (t+nT)}$.
Note that $nT$ can considerably exceed the characteristic time ${\cal T}$
of evaporation of mini-time-machine (which can be smaller than $10^{-28}$ s
for MTM possibly created at LHC and is negligible as compared to life-time
of all ordinary particles), both because $T$ can be much greater
than ${\cal T}$ and because $n$ can be large.

Similar phenomenon will take place with the loss of energy due to
radiation. As shown in Fig.4, the radiated photons can have
different fates -- they can escape from the time machine immediately
or can continue to make time loops in it,-- this does not affect the
radiation rate of original particle. Other effects, related to the
age of particles, like kaon or neutrino oscillations will also be
seemingly accelerated in the presence of MTM.

\subsection{Effective non-locality, possible $CPT$ and unitarity
violation}

As shown in Fig.4, a photon, emitted by a particle when it was
traveling inside the time machine, can escape, but its emission
point will have nothing to do with that of the escaping particle itself
(if escaping photon was emitted not at the last, but at some
intermediate travel of the particle along the time loop, and if
we notice that trajectories of these travels in space can differ
from time to time -- as shown in Fig.4).
This means that from the point of view of external observer
a kind of non-local emission of photon took place: a potentially
non-local effective interaction is generated by time machine.
This opens a room for $CPT$-violation, though more detailed
analysis is needed to decide whether it can indeed be caused by
{\mtm}\ (see also \cite{HawH} for a toy two-dimensional model discussion
of the issue)\footnote{For possible CPT violation effects due to non-trivial topology
see, e.g., a recent paper \cite{Klin} and references therein.}.

Naive unitarity violation in time-machine geometries
is considered very probable \cite{unit}, because of
existence of the Jinnee-type world histories \cite{LN}, shown in Fig.5,
which are not fully controlled by initial conditions at remote past
and should be somehow averaged over.
Of course, after evaporation of MTM the {\it formal} unitarity
should be recovered,
like it happens in the case of evaporating black holes \cite{evabh}
(or molecular theory).
However, physically relevant thermalization-style effects
remain and can probably be observed.


\begin{figure}
\epsfxsize 500pt
\begin{center}
\epsffile{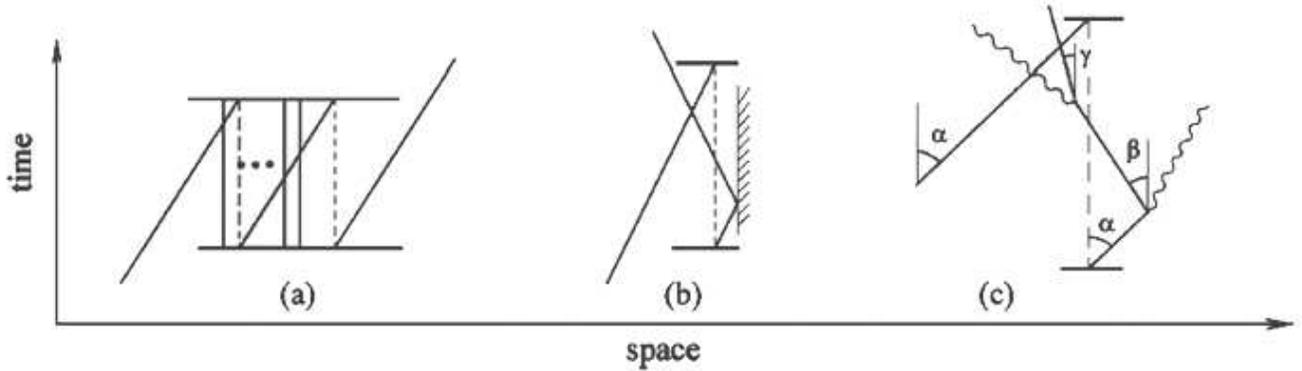}
\end{center}
\caption{\footnotesize Encounters of external particle with
the ``residents" of the time machine,
nicknamed "Jinn" in ref.\cite{LN}. They can interfere with
the incoming and outgoing particles and are well observable, but
they are not controlled by specification of {\it in} or {\it out}
states, what requires a more accurate formulation of unitarity
when particle theory is considered in the time-machine geometries.
{\bf (a)} Vertical lines (classically unobservable Jinn-at-rest)
can be added in any number at any place of
the cuts. This represents the ambiguity in the world-history of a
particle traveling through the time machine. In the picture it is
assumed that particles interact as in central collisions of tiny balls:
they exchange momentum and energy in every act of interaction.
{\bf (b)} A non-trivial meeting with Jinnee.
Encounter with the Jinnee is not controlled by the initial
condition at remote past.
One and the same pattern (which contributes exactly once to
the functional integral) allows different {\it interpretations}:
one can say that a particle gets through the time-machine without
interaction with the Jinnee living inside or that a particle itself
passes along a time-loop inside time machine and escapes, with
changes in its motion, caused by self-interaction.
{\bf (c)} Here the aging (dissipating) Jinnee is shown: as time goes,
our particle can radiate (or decay or dissipate energy in other ways).
The remaining kinetic energy and thus the age of the particle is
characterized by the angle between the world line and time-axis.
It is clear from the picture that as result of encounter with the
Jinnee the particle becomes older -- in accordance with the second
interpretation in (b): that it have spent extra proper time inside
the time machine.
The typical (proper-time) history of a "wild" Jinnee is as follows
(see \cite{LN} for discussion of artificially created Jinn).
Radiation can cause incoming particle to be trapped inside the
time machine and become a Jinnee. After that the aging Jinnee looses
energy and becomes sterile Jinnee-at-rest, which has no more energy
to loose and can no longer radiate. Still, since the energy of
resting Jinnee can exceed its mass due to zero-fluctuation effects
in the time-machine of a finite size, the quantum
interaction with virtual pairs can allow the Jinnee to escape:
this effect contributes to quantum evaporation of time-machines
(gravitational fields, if any, can also be considered as a sort of
Jinn).
}
\end{figure}

\subsection{Collective effects}

Fig.4 shows that a time-machine converts, at some stages of
evolution, a single particle into an ensemble of co-existing
particles. Moreover, the particles in the ensemble are copies and
thus are strongly correlated. This opens a room for various
collective effects to occur, if interaction is taken into account,
up to bose-condensation, superfluidity and superconductivity of
copies inside the time-machine. Since phase transitions may affect
the behavior of particles, including, say, their decay rates and
radiation distributions, they may affect the outcome of particle
collisions. This kind of collective effects, caused by cloning of a
single particle, seems to be absolutely peculiar for time machines.

\subsection{Extra production of antiparticles at intermediate stage}

As clear from the presentation of wormhole geometry in the right-hand-parts
of Figs.1 and 4, while traveling through the time-machine, a probe particle
will sometime move
backwards in time from the point of view of the laboratory frame.
Probably, at this part of its history it can be considered as
antiparticle. In other words, the {\mtm}\ seem to temporarily
convert particles into antiparticles, however, as clear from Fig.6,
for external observer the decay products of {\it these} antiparticles are
indistinguishable from those of original particles, at least in the
theory without CP-violation. Thus, experimental consequences of temporal
presence of antiparticles are obscure.

\begin{figure}[ht]
\epsfxsize 400pt
\begin{center}
\epsffile{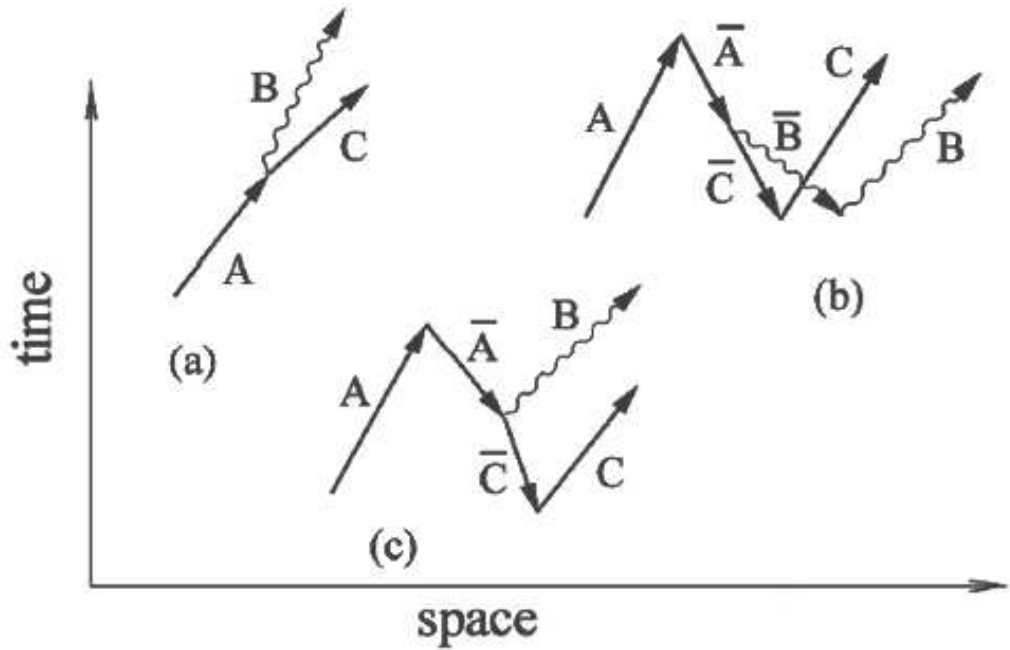}
\end{center}
\caption{\footnotesize
Irrespective of time-directions in which the particle $A$
can move inside a time-machine (and its possible conversion into
anti-particle $\bar A$), the observer at remote future will see
only $B$ and $C$ (and not $\bar B$ or $\bar C$) in the products
of $A$ decay.
{\bf (a)} A particle $A$ can emit $B$ and turn into $C$.
{\bf (b)} When moving backwards in time, $C$ looks like a normally moving
anti-particle $\bar C$, which turns into $\bar A$ after collision with
$\bar B$.
{\bf (c)} Alternatively, $\bar C$ can convert into $\bar A$ by emission of
$B$.}
\end{figure}

If $CP$ is violated, a more interesting
question arises: antiparticles should have 
coupling constants, which are complex conjugate of their particle's
counterparts ($CP$-violating mass terms, e.g.
$i\mu(K_0^2-\bar K_0^2)$ as opposed to the ordinary $m^2K_0\bar K_0$,
are usually considered as additional valence-two vertices in
Feynman diagrams).
In the first quantization formalism this means that
when moving backwards in time the particle should have different
parameters. Such abrupt change is not a problem in the formalism,
involving singular paths of Fig.3c, where world lines of
particles and antiparticles lie inside light cones and are well
separated. However, adequate description in terms of
smoothly $U$-turning paths of
Fig.3b is less obvious. In Fig.4, associated with the time-machine
geometry, the light-cone is {\it smoothly}
turning upside down along the world line, what resembles the
``geometric" Fig.3b, rather than ``physical" Fig.3c.
All this means that the story
of geometrical first-quantized formulation of theories with
CP-violation and, in particular, the problem of
particle-antiparticle conversion in MTM is somewhat obscure and
deserves further investigation.

\section{Conclusion}

No question, the statements in the previous sections and even their mutual
compatibility are disputable or at
least require a lot of comments and justification.
However, attempts to make above naive arguments more rigorous would take
us back to controversial discussion of time-machine physics, which we
suggested to avoid in the Introduction.
Instead, as we suggested, one can try to make equally naive estimates
of the values of above effects, assuming that -- for one or another reason --
they can avoid the existing or the future counter-arguments and still show up
in physical experiments. 
Of principal importance for reliable estimates is deeper understanding
of the issues, mentioned in s.\ref{s2}.

\section*{Note added}

 This paper concerns a subject which lies
at the intersection of two controversial subjects: time-machines in
general relativity and low-energy quantum gravity. Recent paper
\cite{AV} raises the same kind of questions, and we decided to put
this old paper to the ArXiv in order to contribute to the
discussion. References have not been updated since 2005 and we
apologize for non-indicating many new contributions to particular
subjects mentioned in the text.

\section*{Acknowledgements}

We are grateful to T.Mironova for help in making figures.
This work was supported in part by the EU under the RTN contract
MRTN-CT-2004-512194.
The A.M.s acknowledge the support of two NATO travel grants and
the hospitality of the Department of Physics of the University of Crete, where
this work was done.
Our work is partly supported by the Federal Program of the
Russian Ministry of Industry, Science and Technology No
40.052.1.1.1112, by Russian Federal Agency of Atomic Enery, by the grants RFBR
04-02-16538a (Mironov), RFBR 04-02-16880 (Morozov), by the Grant of
Support for the Scientific Schools 8004.2006.2, NWO project
047.011.2004.026, INTAS grant 05-1000008-7865
and ANR-05-BLAN-0029-01 project.

\end{document}